\documentclass[amssymb,prd,superscriptaddress,aps,nofootinbib,showpacs]{revtex4-1}
\usepackage{graphicx, epsfig, amssymb} %include figure files
\usepackage{amsmath, amsfonts}
\usepackage{bm} %include bold math: \bm{} creates bold letters in math mode
\usepackage[breaklinks]{hyperref}
\usepackage{color}
 \usepackage[usenames]{xcolor}

\newcommand{\be}{\begin{equation}}
\newcommand{\ee}{\end{equation}}                  
\newcommand{\bea}{\begin{eqnarray}}
\newcommand{\eea}{\end{eqnarray}}

\begin{document}
%%%%%%%%%%%%%%%%%%%%%%%%%%%%%%%%%%%%%%%%%%%%%%%%%%%%%%%%%%%%%%%%%%%
 
\title{Van Der Waals Black Holes in $d$ dimensions}

 \author{T\'erence Delsate}
 \email{terence.delsate@umons.ac.be}
 \affiliation{Department of Theoretical and Mathematical Physics,
 University of Mons, 7000 Mons, Belgium}

 \author{Robert Mann}
 \email{rbmann@sciborg.uwaterloo.ca}
 \affiliation{Department of Physics and Astronomy, University of Waterloo, Waterloo, Ontario, Canada, N2L 3G1}

\date{\today}

\begin{abstract}
We generalize the recent solution proposed by Rajagopal et al. to arbitrary number of dimensions and horizon topologies. We comment on the regime of validity of these solution. Among our main results, we argue that the Van Der Waals (VDW) black hole (BH) metric is to be interpreted as a near horizon metric. This is supported by inspecting the energy conditions. We analyze the limiting cases of a perfect fluid, interacting points and non interacting balls gas equation of state and map them to known black holes. Finally, we provide a case study by comparing the Reissner-Nordstr\"om and VDW BH close to the horizon and show that they are qualitatively similar for some range of the horizon radius. 
\end{abstract}

%04.50.Gh 	Higher-dimensional black holes, black strings, and related objects
%04.20.Jb: Exact solutions
% 04.70.Bw 	Classical black holes
% 04.70.Dy 	Quantum aspects of black holes, evaporation, thermodynamics
%02.30.Jr: Partial differential equations
%04.25.-g: Approximation methods; EoMs
%04.40.Dg: Relativistic stars: structure, stability, and oscillations 
%04.50.Kd: Modified theories of gravity
%04.60.Cf: Gravitational aspects of string theory
%97.60.Jd: Neutron stars
%11.10.Gh: Renormalization

\maketitle

\tableofcontents

%%%%%%%%%%%%%%%%%%%%%%%%%%%%%%%%%%%%%%%%%%%%%%%%%%%%%%%%%%%%%%%%%%%
\section{Introduction}
%%%%%%%%%%%%%%%%%%%%%%%%%%%%%%%%%%%%%%%%%%%%%%%%%%%%%%%%%%%%%%%%%%%
The thermodynamic properties of black holes provide important clues to the behaviour of quantum gravity. The physics of asymptotically AdS black holes has been of considerable interest for quite some time \cite{GibbonsHawking:1977} because of the  AdS/CFT correspondence, which led to an effort to understand strongly coupled thermal field theories living on the AdS boundary.   
Recently the thermodynamic properties of these kinds of black holes have been a subject of intense interest because of their 
qualitatively similar behaviour to a Van der Waals fluid  once rotation and/or charge are added \cite{ChamblinEtal:1999a,Cvetic:1999ne,Niu:2011tb}. However the analogy is complete \cite{KubiznakMann:2012,Gunasekaran:2012dq} only when the the cosmological constant $\Lambda <0$ is interpreted as thermodynamic pressure $P=P_\Lambda$, \cite{CreightonMann:1995,CaldarelliEtal:2000}
\be
P_\Lambda=-\frac{\Lambda}{8\pi}=\frac{3}{8\pi l^2}\,,
\ee
 and is allowed to vary in the first law of black hole thermodynamics, 
\be
\delta M=T \delta S+V \delta P +\dots,
\label{first}
\ee
where the quantity $V$ 
\be
V=\bigg(\frac{\partial M}{\partial P}\bigg)_{S,\dots}\,.
\label{V}
\ee
is the thermodynamic volume and is conjugate to $P$ \cite{KastorEtal:2009, Dolan:2010}.  
From this, an equation of state 
$P=P(V,T)$ can be written down for a given black hole.  By identifying  the black hole and fluid temperatures $T\sim T_f$,  volumes $V\sim V_f$, and  pressures $P\sim P_f$ a proper mapping to the  corresponding fluid equation of state can be made \cite{KubiznakMann:2012,Gunasekaran:2012dq}. Charged and/or rotating AdS black hole thermodynamics has been shown to qualitatively mimic the behaviour of a standard Van der Waals (VdW) fluid.  The VdW liquid/gas phase transition
corresponds to a small/large black hole first-order phase transition, which terminates at a critical point characterized by the standard mean field theory critical exponents. The  Gibbs free energy exhibits a  swallowtail catastrophe in both cases 
\cite{KubiznakMann:2012, Gunasekaran:2012dq}.  This analogy extends to a broad class of black holes in higher dimensions \cite{Altamirano:2014tva}.

A VdW fluid is described by the VdW equation of state (EOS)
\be
T=\Bigl(P+\frac{a}{v^2}\Bigr)(v-b) 
\label{VdW}
\ee
 which is a closed form 2-parameter equation of state, valid in any spatial dimension \cite{Gunasekaran:2012dq}.  Here  $v=V/N$ denotes the specific volume of the fluid,  with $N$ counting the degrees of freedom of the fluid. The parameter $a>0$ is a measure of the intermolecular attraction between the fluid constituents and the parameter $b$ is a measures of their volume. However the corresponding equations of state   
black hole and the VdW fluid, though qualitatively mathematically similar, are not identical.  Consequently neither charge $Q$ nor rotation $J$  (or other parameters for more complicated black holes \cite{Altamirano:2014tva}) can be directly identified with the fluid parameters  $a$ and $b$.  In the following, we will denote the pressure associated with the cosmogical constant $P_\Lambda$ to avoid confusion with the pressure of the fluid sourcing the Einstein equations.
 
Recently a proposal was put forward to construct an exact black hole analog of the VdW fluid in 4 space-time dimensions\cite{Rajagopal:2014ewa}, inspired by this qualitative analogy.  There is a fluid yielding a metric from the Einstein equations that can indeed give rise to the equation of state (\ref{VdW}).  However it can only obey the standard energy conditions for sufficiently small pressures and in a region sufficiently close to the event horizon.
Here we extend this result to arbitrary number of dimension and arbitrary horizon topologies. The higher-dimensional cases may be of some physical relevance;
in particular, in 5 dimensions $AdS$ black holes have a well understood 4 dimensional $CFT$ counterpart. We further examine in more detail properties of these black holes such as their domain of existence, and interpret the free constants appearing in the solutions. We also discuss the limiting cases of non interacting gas with finite size, interacting points and perfect gas, and map them to the corresponding black hole solutions.

This paper is organized as follows: In Sec. \ref{sec:AdSBH}, we review the basic thermodynamic properties of $AdS$ black holes in $d$ dimensions and extend them to  a parametrized metric suited for our construction \cite{Rajagopal:2014ewa}. In Sec. \ref{sec:VDWBH}, we derive the $d$-dimensional solution describing a VDW black hole, and discuss the domain of existence of the latter. Next, we analyze the source of the Einstein equation yielding the VDW black hole and discuss their energy conditions. In Sec. \ref{sec:cons} we provide an explicit realization of a VDW black hole satisfying the energy conditions in a range close to the horizon. In Sec. \ref{sec:limit}, we discuss the limiting cases where $a=0$ or $b=0$. Finally, Sec. \ref{sec:case} is devoted to providing an explicit comparison of the Reissner-Nordstr\"om black hole with the VDW black hole, in order to understand the qualitative similarities from a metric point of view. We summarize our finding in the concluding section \ref{sec:concl}.

\section{$AdS$ Black Hole in $d$ dimensions}
\label{sec:AdSBH}
\subsection{Metric ansatz and matter source}

Let us first review the basic properties of the $AdS$ vacuum black hole solution in $d$ dimensions, described by the  $AdS$-Tangherlini black hole \cite{Tangherlini:1963}:
\be
ds^2 = -f(r) dt^2 + \frac{dr^2}{f(r)} + r^2 d\Omega_{k,d-2}^2,
\ee
where $d\Omega_{k,d-2}$ is the line element on the unit $d-2$ constant curvature surface, with  $k$ the sign of the curvature. The vacuum solution is given by
\be
f(r) = \frac{r^2}{\ell^2 } + k - \frac{\mu}{r^{d-3}},
\ee
where $\ell$ is the cosmological radius defined by $\Lambda = -\frac{(d-1)(d-2)}{2\ell^2}$ and $\mu$ is related to the AdM mass $M$ of the black hole:
\be
M = \frac{(d-2)\Omega_{k,d-2}}{16\pi}\mu.
\label{eq:Madm}
\ee
 
This solution satisfies the $AdS$ vacuum equation $G_{ab} +\Lambda g_{ab} = 0$. In the following, we will follow the construction of \cite{Rajagopal:2014ewa} and build $f(r)$ 
such that a desired equation of state on the black hole thermodynamics is obtained. Generically, the solution of this procedure will no longer be a vacuum solution but will 
contain sources. A satisfying solution will be obtained if some basic properties remain valid, such as e.g. preservation of energy conditions.

Generically, given a function $f$, the stress tensor is obtained using the field equations. Here, we assume the stress tensor to be given by an  anisotropic fluid of the form
\be
T^{ab} = \rho e_0^a e_0^b + \sum_i P_i e_i^a e_i^b = G^{ab} + \Lambda g^{ab},
\ee
where $e_\mu^a$ are the components of the vielbein and $i=1,\ldots,d$ and where 
\bea
&&\rho = -P_r =-\frac{(d-2) f'(r)}{2 r}-\frac{(d-3) (d-2) f(r)}{2 r^2}+\frac{(d-3) (d-2) k}{2 r^2}+ 8\pi P_\Lambda, \nonumber\\
&&P_\alpha = \frac{(d-3) f'(r)}{r}+\frac{(d-4) (d-3) f(r)}{2 r^2}-\frac{(d-4) (d-3)k}{2 r^2}+\frac{f''(r)}{2}-8\pi P_\Lambda,
\label{eq:stress}
\eea
where $\alpha$ labels the coordinates in the angular sector.  Note that we use the convention $G =1/(8\pi)$, while \cite{Rajagopal:2014ewa} uses $G=1$, $G$ being Newton's constant.

%%%%%%%%%%%%%%%%%%%%%%%%%%%%%%%%%%%%%%%
\subsection{Thermodynamic properties}
%%%%%%%%%%%%%%%%%%%%%%%%%%%%%%%%%%%%%%%

In $d$ space-time dimensions the cosmological constant will be interpreted as thermodynamic pressure, given by \cite{CreightonMann:1995,CaldarelliEtal:2000,KastorEtal:2009, Dolan:2010,Kubiznak:2012wp}
\be
P_\Lambda = -\frac{\Lambda}{8\pi} = \frac{(d-1)(d-2)}{16\pi\ell^2}
\label{pressd}
\ee
and  without loss of generality, we write the function $f$ as
\be
f = \frac{r^2}{\ell^2} - \frac{\mu}{r^{d-3}} - h(r,P_\Lambda),
\label{eq:ansatzf}
\ee
where $h$ is a function to be determined and $\mu$ is related to the mass of the black hole as in \eqref{eq:Madm}.

The temperature $T$ of a black hole in this class of solutions is obtained by requiring regularity of the Euclidean section, yielding
\be
T = \frac{f'(r_h)}{4\pi},
\ee
where $r_h$ is the location of the horizon radius. The entropy is given by the Bekenstein-Hawking law
\be
S = \frac{\mathcal A}{4} = \frac{\Omega_{d-2}}{4}r_h^{d-2},
\label{eq:entropy}
\ee
where $\Omega_{n}$ is the surface of the $n$-dimensional unit sphere
and the mass is given by $M$ provided $h$ doesn't contain terms linear in $1/r^{d-3}$, and can be expressed in terms of other quantities by solving 
$f(r_h) = 0$ for $M$.

The first law of thermodynamics then reads
\be
dM = T dS + V_\Lambda dP_\Lambda,
\ee
where the momentum $V_\Lambda$ conjugated to $P_\Lambda$ is interpreted as a volume, given by
\be
V_\Lambda = \left.\frac{\partial M}{\partial P}\right|_{S},
\ee
and $M$ is the enthalpy,
\be
M = \frac{(d-2)}{16\pi}\Omega_{k,d-2}r_h^{d-3}\left( \frac{16\pi P_\Lambda r_h^2}{(d-1)(d-2)} - h(r_h,P_\Lambda)  \right).
\label{eq:mass}
\ee

The specific volume is then given by
\be
v_\Lambda = \kappa\frac{V}{\mathcal A},
\ee
where $\kappa = 4\frac{d-1}{d-2}$.

\section{Van der Waals Black Holes}
\label{sec:VDWBH}
\subsection{Construction of the solution}

Using \eqref{eq:ansatzf}, the thermodynamic quantities defined in the previous sections reduce to 
\bea
&&T = -\frac{(d-3) h(r_h,P_\Lambda)}{4 \pi r_h}-\frac{h^{(1,0)}(r_h,P_\Lambda)}{4 \pi }+\frac{4 P_\Lambda r_h}{d-2}\label{TL}\\
&&M = \frac{ \Omega_{d-2} }{d-1}P_\Lambda r_h^{d-1}-\frac{\Omega_{d-2}}{16 \pi }(d-2) r_h^{d-3}  h(r_h,P_\Lambda),\label{ML}\\
&& v_\Lambda = \frac{4 r_h}{d-2}-\frac{(d-1) h^{(0,1)}(r_h,P_\Lambda)}{4 \pi r_h} \label{vL}
\eea
and the entropy $S$ is given by \eqref{eq:entropy}. Specifying an equation of state yields a set of  partial differential equations for the function $h$. Here we impose the equation of state (\ref{VdW})
\be
T = \left( P_\Lambda +\frac{a}{v_\Lambda^2}\right)\left( v_\Lambda - b  \right),
\label{eq:VDW}
\ee
where we set $v=v_\Lambda$, $P=P_\Lambda$ and will regard $T$ as being given by (\ref{TL}).  
To solve the resultant partial differential equations that emerge from  \eqref{eq:VDW} requires an ansatz.  Here we employ
$h = A(r) - P_\Lambda B(r) $; inclusion of higher powers of $P_\Lambda$ does not yield solutions consistent with the asymptotic  AdS structure
 \cite{Rajagopal:2014ewa}. Note that strictly speaking, the equation of state provides a boundary condition on the event horizon. However, if this condition is satisfied everywhere, obviously it is also satisfied on the horizon. We shall first follow this strategy and then comment on the limitations of this approach.  

Our ansatz for $h$ yields two ordinary differential equations for $A$ and $B$. 
The equation for $B$ is independent of the number of the number of dimensions and is given by
\be
r B' - 2 B + 4\pi r b = 0,
\ee
which is solved by
\be 
B =   4\pi r (C_\ell r  +  b),
\label{Bsol}
\ee
where $C_\ell$ is an integration constant whose meaning will be discussed later.  

The equation for $A$ is
\bea
&&r_h A'(r_h) + (d-3) A(r_h) 
%\nonumber\\
%&&
+\frac{16 \pi^2  a (d-2) r_h^2 \left((d-2) (d-1)B(r_h)+16\pi r_h^2- 4\pi r_h b (d-2)\right)}{ ((d-2) (d-1)B(r_h)+16\pi r_h^2)^2}
%\nonumber \\
%&&
=0 
\eea

Using (\ref{Bsol}), the general solution of this equation is given by (see Appendix)
\begin{align}\label{Asol}
&A(r;d) = \frac{a \pi x^4(d-2)}{(1+ \tilde C_\ell)(d^2-1)}\left((d-4) \, _2F_1\left(1,d+1;d+2;-x\right)-\, _2F_1\left(2,d+1;d+2;-x\right)\right)\nonumber\\
&\quad -\frac{\pi  a (d-2) x \left((d-4) (d-1) x^2-(d-3) d x+(d-1) d\right)}{(\tilde C_\ell+1) (d-1)^2 d} + \frac{\tilde C_M}{r^{d-3}},
\end{align}
where $x = \frac{4 (\tilde C_\ell+1) r}{b (d-2) (d-1)}$, $\tilde C_\ell = (d-1)(d-2) C_\ell /(16\pi)$, $\tilde C_M$ is a constant of integration and $_2F_1$ is the hypergeometric function.

For $d$ a positive integer, the solution \eqref{Asol} reduces to a finite number of terms.  
For particular values of $d=4,5,6,7$ and arbitrary values of $C_\ell$ we obtain
\bea
&&d=4:
% \nonumber\\
% &&
A = -\frac{4 \pi  a}{r (3 C_\ell+2)^2}\Biggl( 3  \left(-\frac{b^2}{3  C_\ell r+3 b+2 r}+C_\ell r\right)
% \nonumber\\
% &&
% \quad
-4 b \log (4 \pi  (3  (C_\ell r+b)+2 r))+2 r \Biggr) + \frac{C_M}{r}\nonumber\\
&&d=5:
% \nonumber\\
% &&
A=-\frac{3 \pi  a}{4 r^2 (3  C_\ell+1)^3}\Biggl(\frac{27 b^3}{3  C_\ell r+3 b+r}
% \nonumber\\
% &&\quad
+54 b^2 \log (4 \pi  (3  (C_\ell r+b)+r))-15 b r (3  C_\ell+1)
% \nonumber\\
% &&\quad
+2 (3  C_\ell r+r)^2  \Biggr) + \frac{C_M}{r^2},\nonumber\\
&&d=6:
% \nonumber\\
% &&
A=-\frac{4 \pi  a}{3 r^3 (5  C_\ell+1)^4}\Biggl( -\frac{375 b^4}{5  C_\ell r+5 b+r}
% \nonumber\\
% &&
% \quad
-600 b^3 \log (4 \pi  (5  (C_\ell r+b)+r))+105 b^2 r (5  C_\ell+1)
 \nonumber\\
&&
 \quad\quad
-9 b (5  C_\ell r+r)^2+(5  C_\ell r+r)^3   \Biggr) + \frac{C_M}{r^3}\nonumber\\
&&d=7:
% \nonumber\\
% &&
A=-\frac{5 \pi  a}{6 r^4 (15 C_\ell+2)^5}\Biggl( \frac{1518750 b^5}{15  (C_\ell r+b)+2 r}
% \nonumber\\
% &&\quad
+1012500 b^4 \log (4 \pi  (15  (C_\ell r+b)+2 r))
 \nonumber\\
 &&\quad\quad
-60750 b^3 r (15 C_\ell+2)+1800 b^2 r^2 (15 C_\ell+2)^2
%  \nonumber\\
%  &&\quad
+3 r^4 (15 C_\ell+2)^4-70 b r^3(15 C_\ell+2)^3  \Biggl) 
\label{eq:solvdw}
\eea
where we have retained the constants of integration in the solutions, and $C_M$ is related to $\tilde C_M$ by $d$-dependent transformations that can be computed straightforwardly.

The resultant space-time metric is given by 
\be
ds^2 = -\left( \frac{r^2}{\ell^2} - \frac{\mu}{r^{d-3}} - A(r;d) + P_\Lambda B(r)  \right) dt^2 + \frac{dr^2}{\left(\frac{r^2}{\ell^2} - \frac{\mu}{r^{d-3}} - A(r;d) + P_\Lambda B(r) \right)} + r^2 d\Omega_{k,d-2}^2.
\label{VdWmet}
\ee
Results from \cite{Rajagopal:2014ewa} are recovered in the $d=4$ case by setting $C_\ell=0,\ C_M = 3 ab \pi$.

% 
% \bea
% &&d=4:\; A=-\frac{6 \pi  a b' (b'+x)}{x (3 b'+2 x)}+2 \pi  a \left(\frac{2 b' \log (3 b'+2 x)}{x}-1\right),\nonumber\\
% &&d=5:\; A= \frac{3 \pi  a \left(-27 b'^3+45 b'^2 x+9 b' x^2-2 x^3\right)}{4 x^2 (3 b'+x)}\nonumber\\
% &&\quad-\frac{81 \pi  a b'^2 \log (3 b'+x)}{2 x^2}\nonumber\\
% &&d=6:\; A=\frac{800 \pi  a b'^3 \log (5 b'+x)}{x^3}\nonumber\\
% &&\quad+\frac{4}{3} \pi  a \left(\frac{3 b' \left(\frac{125 b'^3}{5 b'+x}-35 b' x+3 x^2\right)}{x^3}-1\right)\nonumber\\
% &&d=7:\; A=-\frac{421875 \pi  a b'^4 \log (15 b'+2 x)}{16 x^4}\nonumber\\
% &&\quad-\frac{5 \pi  a }{96 x^4 (15 b'+2 x)}\Bigl(759375 b'^5-911250 b'^4 x\nonumber\\
% &&\quad-67500 b'^3 x^2+3000 b'^2 x^3-200 b' x^4+48 x^5\Bigr),
% \label{eq:solvdw}
% \eea
% where $b' = b/r_0,\ x = r/r_0$ and $r_0$  is the integration constant.

\subsection{Interpretation of the constants and domain of existence}
\label{sec:exist}

%The corresponding generalization for $d>4$ is obtained by setting $C_\ell=0,\ C_M = 0$. 

The choice of integration constants is governed by physical criteria \cite{Rajagopal:2014ewa}.  If the cosmological pressure \eqref{pressd} is required to be generated entirely by the fluid, then $C_\ell = 0$.  However if this criterion is relaxed, then the effective AdS length  is  $\ell^{-2}_{eff} = \ell^{-2}(1 + (d-1)(d-2) C_\ell/4) $.  Physically this means that the thermodynamic pressure is given entirely by the cosmological constant, but the fluid has an additional density and pressure that is similar to the cosmic vacuum.  The constant $C_M$, however, simplify yields a shift in the parameter $\mu$; we can therefore set it to zero without loss of generality.
%Likewise, if the mass term is generated entirely by the fluid parameters $(a,b)$, then $C_M=0$.  However this criterion can also be relaxed.

The constant $a$ can be set so that the constant term in $r$ in the asymptotic form of $f$ is $k$, similar to the $d=4$ case.  This leads to $a = \frac{d-3}{\pi  (d-2)}k$.  This means that the intermolecular attraction of the fluid is governed by the curvature of event horizon.  A VdW black hole with a flat horizon section has $a=0$, and one with constant negative curvature (corresponding to topological black holes \cite{Aminneborg:1996iz,Mann:1997,Mann:1998} ) has an effective intermolecular repulsion.

We note that the Van der Waals black holes \eqref{VdWmet} do not possess horizons for all generic values of the parameters. Indeed, for some values of the parameters, the horizon radius crosses $0$ and becomes negative.  Such solutions correspond to solitons (or bubbles); we shall not consider them here.

The allowed region in parameter space for black holes is given by $r_h > 0$.  The boundary of this region can be obtained by solving for the mass in terms of other parameters upon  imposing $r_h = 0$.  This leads to 
\bea
\frac{16 \pi M}{2\Omega_2}&\geq&-\frac{4\pi a b  (1+4 \log (12 \pi  b))}{(2+3  C_\ell)^2}   ,\ \mbox{ for } d=4,\\
\frac{16 \pi M}{3\Omega_3}&\geq& \frac{27\pi  a b^2  (1+6 \log (12 \pi  b))}{4(3  C_\ell+1)^3}   ,\ \mbox{ for } d=5,\nonumber\\
\frac{16 \pi M}{4\Omega_4}&\geq&-\frac{100 \pi  a b^3  (1+8 \log (20 \pi  b))}{(5  C_\ell+1)^4}  ,\ \mbox{ for } d=6,\nonumber\\
\frac{16 \pi M}{5\Omega_4}&\geq&\frac{84375 \pi  a b^4 (1+10 \log (60 \pi  b))}{(15  C_\ell+2)^5}  ,\ \mbox{ for } d=7,\nonumber
\eea
and other numbers of dimensions are straightforwardly computed. 

Note that the specific volume (\ref{vL}) becomes
\be\label{vspec}
v = \left(\frac{4}{d-2}+ (d-1)C_\ell\right) r_h + (d-1)b
\ee
yielding 
\be\label{eq:ipe-ratio}
\mathcal{R} = \left(\frac{(d-1) {V}}{\Omega_{d-2}}\right)^{\frac{1}{d-1}}\left(\frac{\Omega_{d-2}}{ {{\mathcal A}}}\right)^{\frac{1}{d-2}}
= \left(1+ \frac{(d-1)(d-2)}{4}\left(C_\ell + \frac{b}{r_h}\right)\right)^{\frac{1}{d-1}}
\ee
for the isoperimetric ratio $\mathcal{R}$.  This will obey the reverse isoperimetric inequality \cite{CveticEtal:2010} provided
$C_\ell r_h + b > 0$, or in other words if $C_\ell$ is not too negative.

Finally, we  stress that, for a given set of parameters (and horizon radius), the solution \eqref{VdWmet} derived here need only hold in a neighbourhood of the horizon, since the relation \eqref{eq:VDW} involves quantities defined at the horizon. The highest derivative order appearing in this relation is $1$. Therefore, the same thermodynamic properties are satisfied  in a neighborhood of the horizon by black holes with 
\be
f(r) = 1-\frac{\mu}{r^{d-3}} -h(r,P_\Lambda) \mathcal H(r,P_\Lambda),
\ee
where $\mathcal H$ is an arbitrary function with the following properties at the horizon
\be
\mathcal H(r_h,P_{\Lambda}) = 1,\ \mathcal \partial_r \mathcal H(r_h,P_\Lambda) = 0 = \partial_{P_\Lambda} \mathcal H(r_h,P_\Lambda).
\ee
Indeed, in this case, 
\bea
&&\left.\partial_r (\mathcal H(r,P_{\Lambda}) h(r,P_\Lambda))\right|_{r=r_h} = \left.\partial_r h(r,P_\Lambda)\right|_{r=r_h},\ \left.\partial_{P_\Lambda} (\mathcal H(r,P_{\Lambda}) h(r,P_\Lambda))\right|_{r=r_h} = \left.\partial_{P_\Lambda}( h(r,P_\Lambda)\right|_{r=r_h},\\
&&\left.\mathcal H(r,P_{\Lambda}) h(r,P_\Lambda)\right|_{r=r_h} = h,
\eea
leading to the exact same equation for $h$ as the one obtained with $\mathcal H =1$.

The function $\mathcal H$ can serve as a modulating function for the fluid. If it falls off
 sufficiently fast for large radii, $\mu$ remains unambiguously related to the mass (or enthalpy). This also ameliorates 
 the consequences of  the linear and logarithmic terms in \cite{Rajagopal:2014ewa}. If the falloff rate of $\mathcal H$ is sufficiently rapid, the conditions on the constant of integration $C_\ell,\ C_M$ can  be relaxed (as noted above), and the near horizon spacetime  has an `effective cosmological radius' different from the one given by $\Lambda$.

However, this remark is true only in the case where we assume a given horizon radius, or alternatively a  given mass. If the purpose is to build a family of black holes (parametrized, say, by horizon radius) where all solutions satisfy the VdW EOS, then $\mathcal H =1$.  Furthermore, in the general cases where $\mathcal H \neq 1$ and no restriction is applied to the fall-off of $\mathcal H$, it is no longer clear that $M$ be regarded as enthalpy, and that the volume $v$ will remain the same. This is because in $v$ is not a near horizon property but is instead a global property of the spacetime.

\subsection{Source fluid and energy conditions}
Recall that the metric \eqref{eq:ansatzf} with the solution \eqref{eq:solvdw} does not satisfy the $AdS$ vacuum Einstein equations. Instead the VDW black hole is sourced by a non isotropic fluid along with \eqref{eq:stress}.
The source stress tensor is conserved by construction, since it is identically the Einstein tensor, which satisfies the Bianchi identities. 

A crucial remark is that  close to the horizon, while $\rho, P_r$ do not depend on $\mathcal H$, $P_\alpha$ does, since the $f''$ term yields a term of the form  $-\frac{1}{2}{\mathcal H}'' h$.

The energy conditions for a non isotropic fluid are given by
\begin{itemize}
 \item Weak: $\rho\geq0,\rho + P_i\geq0$,
 \item Strong: $\rho + \sum_i P_i\geq0,\rho + P_i\geq0$,
 \item Dominant: $\rho - |P_i|\geq0$.
\end{itemize}
In the case of van der Waals black holes, the density and pressures are given by \eqref{eq:stress} together with the solution for $h$.

In the rest of this subsection, we will consider the VdW black hole generalizing the 4 dimensional solution of \cite{Rajagopal:2014ewa} by setting $\mathcal H = 1$.

We first emphasize that the density and pressure depends solely on $h+k$, since their expressions are linear in $f$ and $f+k+h$ is a vacuum solution. As an important consequence, the properties of the sourcing fluid do not depend on the black hole mass, nor on the parameter $C_M$ since it accounts for a mass term.

We also note that since $C_M$ can always be reabsorbed in the definition of the mass, we set it to $0$ for definiteness. 
Let us analyze whether it is possible to build a VdW black hole that satisfies the energy conditions close to the horizon. 

First, it is instructive to inspect  the small and large $r$ behaviour of the fluid density and pressure.
%%%%%%%%%%%%%%%%%%%%%%%%%%%%%%%%%
For large $r$,  $\rho,\ P_\alpha$ behave as
\bea
&&\rho =-2 \pi   C_\ell d_1 d_2 P_\Lambda-\frac{2 \pi  b d_2^2 P_\Lambda}{r} +\frac{d_3d_2 k-\frac{4\pi  a d_2^2}{ C_\ell d_2 d_1+4}}{2 r^2} \label{rhod}\\
&&P_\alpha = 2 \pi  C_\ell d_2d_1 P_\Lambda+\frac{2 \pi  b d_3d_2 P_\Lambda}{r} + \frac{\frac{4\pi  a d_4d_2}{ C_\ell d_2 d_1+4}-  d_4d_3 k}{2 r^2}\label{Pd}
\eea
where $d_j \equiv (d-j)$.  For small radii, it can easily be checked that
\be
\rho= \frac{d_2d_3}{2r^2} + \mathcal{O}(r)^{-1},\ P_\alpha = -\frac{d_3d_4}{2r^2} + \mathcal{O}(r)^{-1}.
\ee
% 
% 
% \tcr{\bf [I don't think that the small-$r$ expansion is physically meaningful because it is inside the event horizon.  Shouldn't we expand  $\rho$ and $P+\rho$ near the horizon? ]}
%  \tcb{\bf [I think the point is that the fluid doesn't know  a priori about the horizon. I understand the model this way: I can choose the mass, and the constant of integration, this gives me a horizon radius. OR I choose the horizon radius and the Csts, this gives me the mass. But in the density and pressures, the mass never appears, nor does the $C_M$ const. So the fluid doesn't know where the horizon is. Indeed I can end up with the same expressions for density and pressure, but for any horizon radius I want, as long as $C_\ell$ remains the same. I've added a comment on that]}
%  \tcr{\bf [I don't agree with the above.  The portion of the fluid inside the horizon will not be static -- it will collapse toward the singularity. So it is not clear (at least to me) that small-$r$ means anything.]} 

\begin{figure}
 \centering
 \includegraphics[scale=.6]{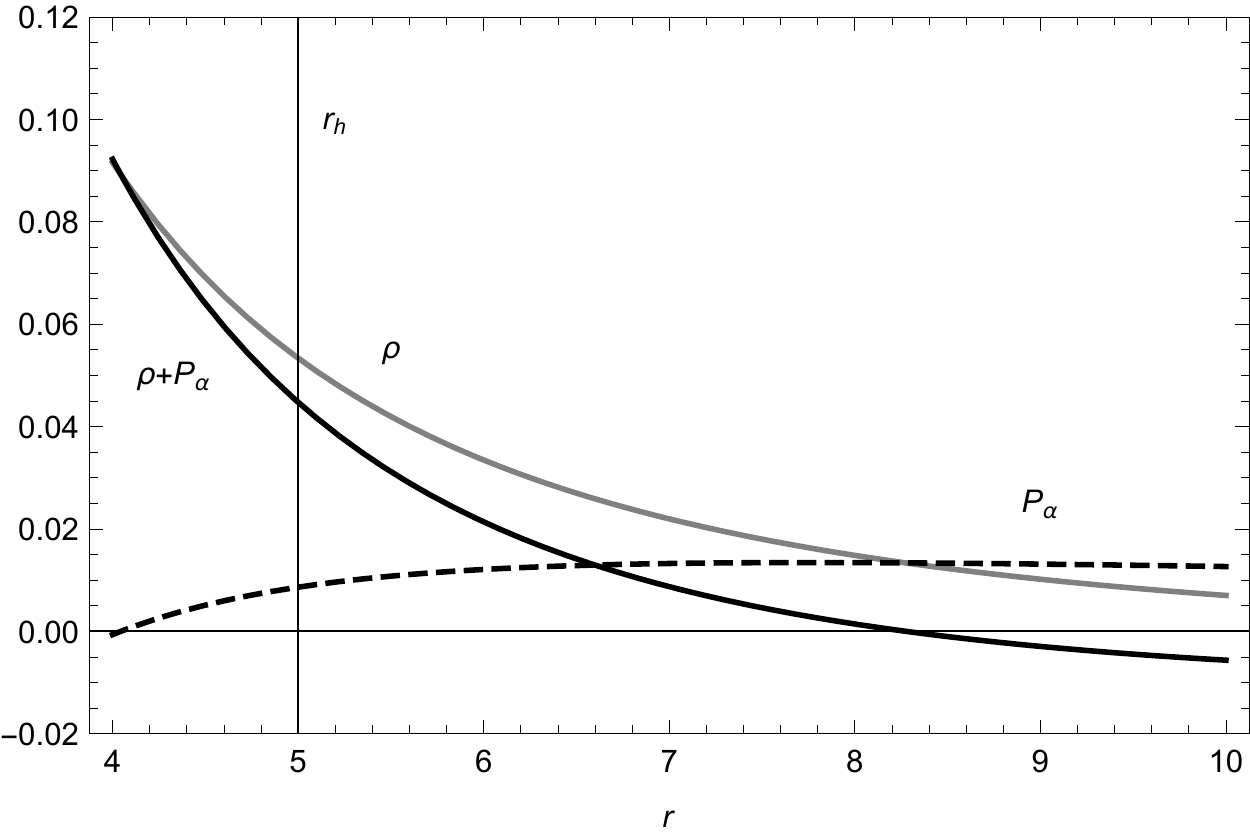}
 \caption{The weak ($\rho,\ \rho+P_\alpha >0$) and strong ($\rho + P_\alpha,\ P_\alpha>0$) energy conditions  for $b=r_h,\ a=2/(3\pi),\ C_\ell=0,\ C_M=0$, $P=0.0009$ and $d=5$. There exists a region where the three functions are positive outside the horizon (located at $r_h=5$). This choice of parameters leads to a positive mass.
 } 
 \label{fig:EC}
\end{figure}

As a consequence, the VdW BH with $C_\ell=0$ \emph{always} violates the weak and strong energy conditions at large radii, and are satisfied at small radii. Although the physical relevance of the small-$r$ is dubious, mathematically 
it follows that there always exists a region where these energy conditions are satisfied. This is illustrated on figure \ref{fig:EC}, where we show the weak and strong energy conditions for ${b=r_h},\ a=1/(2\pi)$ and $d=4$. Other values of $d$ leads to the same pattern. 

Since the source of the stress tensor does not depend on the mass of the black hole, 
%nor the $C_M$ integration constant, 
we conclude that there always exists a parameter set for which the energy conditions can be satisfied \emph{close} to the horizon.  However not all horizon radii lead to positive $M$, see section \ref{sec:exist}. The question then reduces to finding parameter sets so that the region where the energy conditions are satisfied is \emph{outside} the horizon.
%Note we've set $C_M=0$ by absorbing it in the definition of the mass. If one releases this assumption, then it is always possible to adjust $C_M$ such that the mass becomes positive for any radius.

% \tcr{\bf [I don't think we can claim this until we compute the expansion near the horizon (as opposed to the small
% $r$ expansion.]}\tcb{\bf I think it actually is meaningful, since what it means is that whatever the horizon, one can find a set of $a,b,C_\ell$ such that the energy conditions are satisfied in the neighborhoud of the horizon. In fact, I don't think we can say much more than small r or large r.}

Note that for negative values of $C_\ell$ it is always possible to satisfy $\rho\geq0$ everywhere. However the second part of the weak energy condition, namely $\rho+P_\alpha$, always yields  a negative falloff for sufficiently large $r$
\be
\rho+P=-\frac{2 \pi  b d_2 P_\Lambda}{r} + \mathcal{O}(r)^{-2},\mbox{ as } r\rightarrow\infty
\ee
contrary to what was originally reported \cite{Rajagopal:2014ewa}
 for the $d=4$ case.

This further supports the idea that the VdW black hole metric should be thought as a near horizon solution.
In the next section, we will provide evidence that it is possible to build a VDW black hole that satisfies the energy conditions at least in a region close to the horizon by explicitly constructing it in some number of dimensions.

\section{Consistent VdW black hole}
\label{sec:cons}
In this section, we propose a systematic prescription for constructing VdW black holes that satisfy the energy conditions close to the horizon. For definiteness, we set $\mathcal{H} =1$ and $C_M=0$ in the rest of this section.

One possible way of investigating the question of consistent VDW black holes, in the sense that they satisfy the energy conditions, is to look at the near horizon solution for these. However this doesn't teach us much because the way we built the solution is such that we express the mass in terms of the horizon radius quantities. In fact, in a near horizon expansion, the free parameters are $A(r_h),\ B(r_h)$, which in turn are related to $C_M,\ C_\ell$. The procedure is then to use the equations provided by the EOS, and derive them iteratively, since they are the only equations from which we've built the solution.

Instead we present an algorithm in order to build a consistent solution near the horizon.
We require that mass  remain positive, while the energy conditions are satisfied near the horizon for some $r>r_h$. Due to equations  (\ref{eq:stress},\ref{eq:mass}) and the fact that $h(r,P_\Lambda)$ is linear in $P_\Lambda$,
all these quantities can be written in the form $X = X_0 + P_\Lambda X_P $:
\be
\mbox{EC : } \mathcal A_0(r) - P_\Lambda \mathcal A_P(r)\geq 0 \qquad  \mbox{Mass : } \mathcal M_0(r_h) + P_\Lambda \mathcal M_P(r_h)\geq 0,
\ee
where  $\mathcal A_0(r),\ \mathcal A_P(r)$ are both positive and increasing with decreasing $r$ provided
$a = \frac{d-3}{\pi  (d-2)}$, $C_\ell=0$, and $k=1$. This leads to a maximum pressure that saturates the energy condition just outside the horizon:
\be
P_{Max}(r_h) = \mathcal A_0(r_h)/\mathcal A_P(r_h)
\ee
and so for a given a set of parameters $(a,\ b,\ C_\ell)$, it is possible to choose $P_\Lambda < P_{Max}(r_h)$ such that $M>0$ for some finite range of $ r > r_h $. This guarantees that all requisite conditions are satisfied at least in a neighbourhood of the horizon.
 We expect this procedure to be valid over a wide range of the parameter space and we leave a more systematic investigation for future considerations.

\begin{figure}
 \includegraphics[scale=.6]{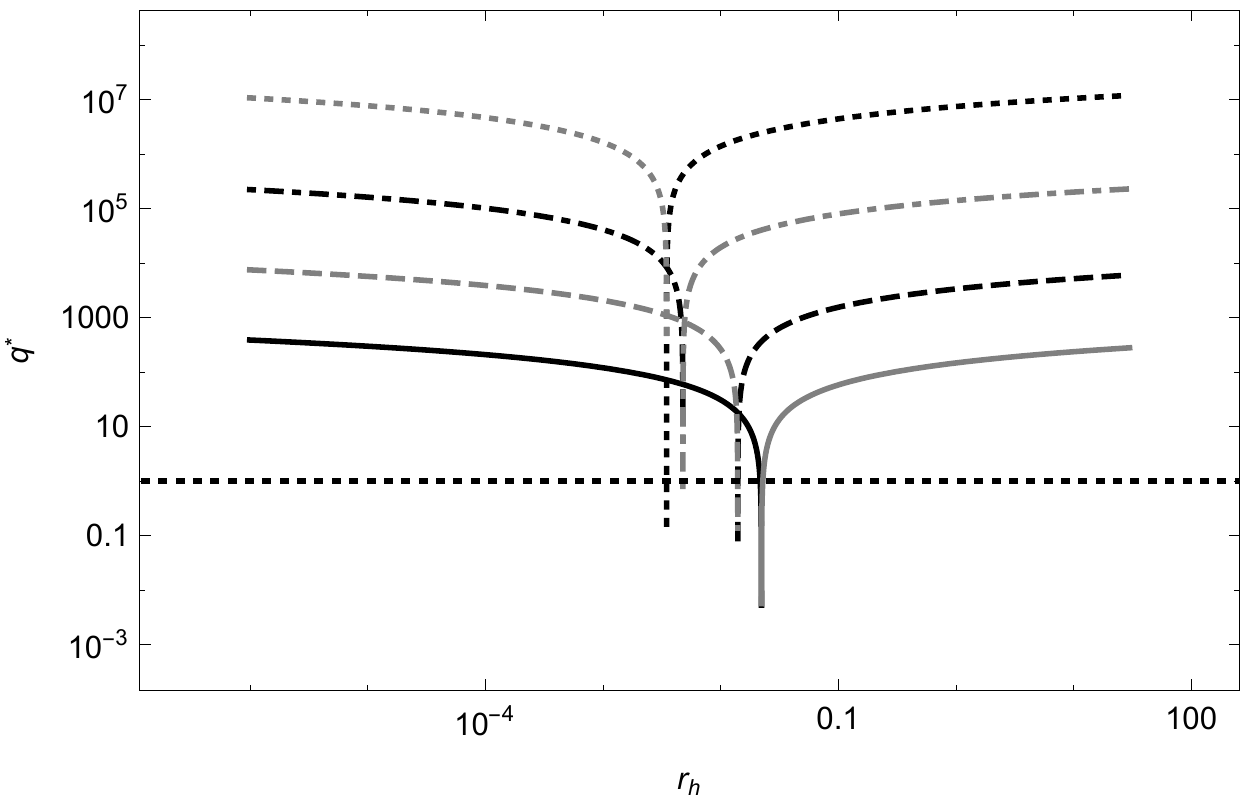}
 \caption{The value of $q^*$ where the mass crosses $0$, and where $P_\Lambda = q^* P_{Max}$. When $q^*\leq1$, the weak energy conditions are satisfied in a neighbourhood of the horizon. If $q^*<0$, then the mass is positive for any choice of $r_h$ and $q^*\leq q\leq 1$. We chose $ b=r_h,\ a=(d-3)/((d-2)\pi),\ C_\ell=C_M=0$. The black parts of the curve are $q^*$ while the gray are $-q^*$ and $d$ ranges from $5$ to $9$, from bottom to top.}
 \label{fig:PQ}
\end{figure}

Setting the pressure a little bit below the maximal value, say $P_\Lambda = q P_{Max},\ q<1$, we can write $M$ as a $q$ independent term and a term linear in $q$, $M = m_0 + q m_1$. The critical value $q^* = -m_0/m_1$  is a function of $r_h$ and corresponds to the fraction of the maximal pressure where the mass changes sign. In particular, when $q^*<1$ the mass is positive and the energy condition considered can be satisfied in a neighborhood of the horizon.
We summarize the behaviour of $q^*$ for a number of dimensions and parameter values in figure \ref{fig:PQ}.

% \tcr{\bf [I went out on a limb -- we have only looked at this for $d=4,5,6,7$.  I am guessing the even/odd trend will hold for all dimensions, but I don't have a proof.  Should we look for one?]}\tcb{\bf[My personal feeling is that it is not necessary, since this regularity can likely be related to the structure of the general solution. I really do expect this trend to hold and would be very surpised if it were not the case. Just for going down from the limb, I've changed a bit the text.. ]}
% 
% \tcb{\bf[I changed the text according to the new results for $b = rh$. This changes the picture. Can you confirm ?]}

We computed $P_{Max}$ as the maximal pressure allowing the weak energy conditions to be fulfilled, as a function of the horizon radius
% When $ q^*\leq 1 $, then there exists a region close to the horizon where the Weak energy conditions is satisfied, since this ensures $M>0$ and a neighborhoud of the horizon where the energy conditions are satisfied.
and systematically found possible to build consistent VDW black holes for the case we considered. We expect that this is possible for all dimensions. 

\begin{figure}
 \includegraphics[scale=.6]{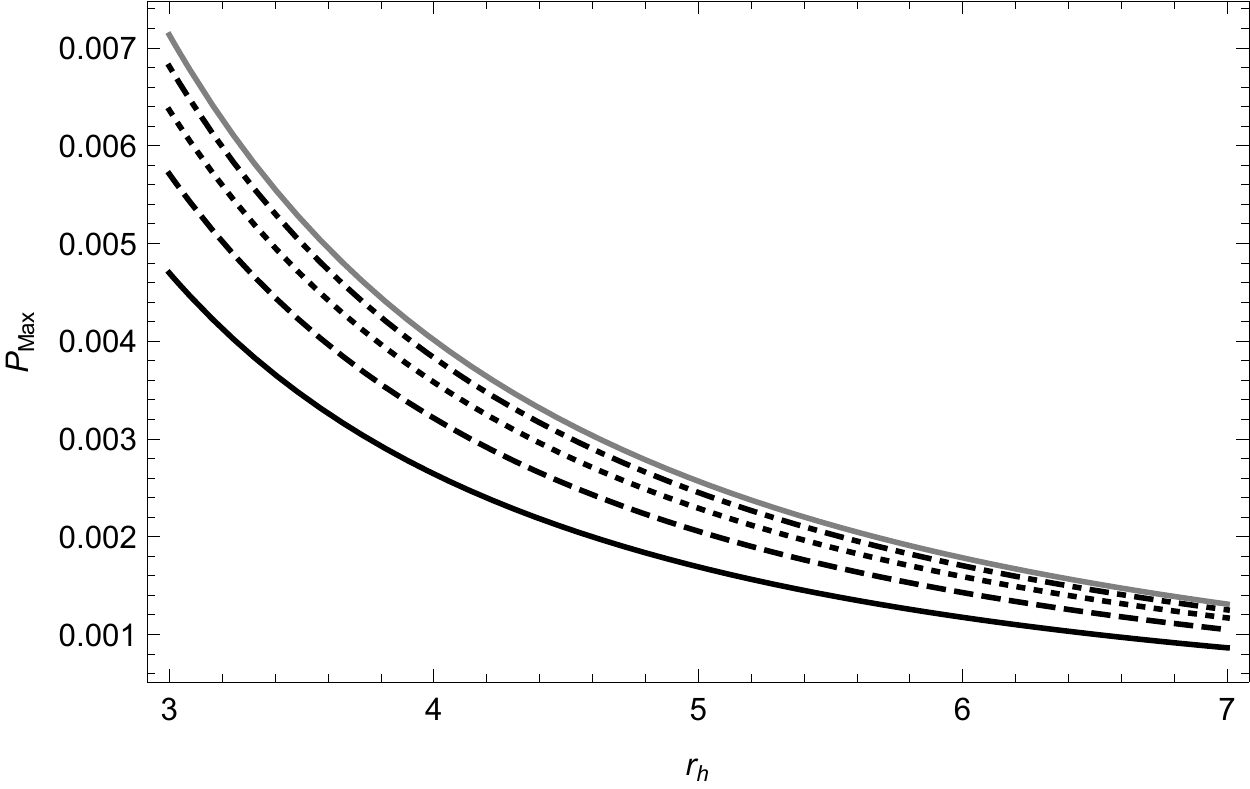}
 \caption{The maximal pressure allowing to satisfy the WEC as a function of the horizon radius, for $b=r_h,\ a=(d-3)/((d-2)\pi),\ C_\ell=C_M=0$ and $d=5,6,7,8,9$ from bottom to top. }
 \label{fig:PMW}
\end{figure}

Note that $b$ is a dimensionful parameter, so  we defined $\tilde b =  b/r_h$. We plot the maximum pressure as a function of $r_h$ for a fixed value of $\tilde b$, and the value $q^*$ such that the mass just vanishes in figures \ref{fig:PQ} and \ref{fig:PMW} for $\tilde b=1$.

Focusing on the WEC (and fixing $k=1$ for definiteness), we find that $P_{Max} = \alpha/r_h^2$, where $\alpha$ depends on $d$ and $\tilde b$. We found (for $d=5,6,7,8,9$) that the ratio $0<r_h/\ell_{Min}<1$, where $\ell_{Min}$ is the value of the cosmological length corresponding to $P_{Max}$. For the case we considered, this means that the intermediate or small black hole phase leads to consistent VDW black holes.

% \tcr{\bf [From which energy condition are you getting this plot?  We need to be clear on this.  What we get for $P_{Max}(r_h)$ will depend on what energy condition we choose. When I construct this plot for the WEC, I get the qualitative behaviour of this graph, but not the quantitative behaviour: I find that the $d=4$ and $d=6$ curves intersect the axis at different points than in this graph.  Also why do you write $P_\Lambda = q^* \Lambda$?  What happened to the $8\pi$?
%  ]}\tcb{\bf[I took the minimum of the $P_{Max}$ obtained for WEC and that obtained for SEC. The $\Lambda$ thing is a typo.. I also messed up the value of $a$. I'll redo these plots too.]}
%  

\section{Limiting cases: perfect gas and interating point gas, non interacting ball gas }
\label{sec:limit}
It is instructive to consider the 3 limits of the VDW equation of state, namely, $a=0, b=0, a=b=0$. These correspond respectively to a gas of free particles with finite size (the ball gas), a gas of interacting point particles, and a perfect gas.

The solutions are given by
\bea
&&f_{perfect} = \frac{16 \pi  (c_\ell+1) P r^2}{(d-2) (d-1)}-\frac{16 \pi  (c_M+M) r^{3-d}}{(d-2) \Omega_{d-2}}\\
&&f_{point} = 4\pi b r + f_{perfect},\\
&&f_{ball} = \frac{(d-2)a\pi}{(d-3)(1+c_\ell)}+f_{perfect},
\eea
where $f_{perfect},\ f_{point},\ f_{ball}$ are the solution for the perfect gas, the interacting point particle and the free ball gas respectively, and $c_M,\ c_\ell$ are constants of integration, rescaled from $C_M,\ C_\ell$. 

We note in passing that the Lemos string \cite{Lemos:1994xp} has the thermodynamic properties of a perfect gas, and the Schwarzschild $AdS$ solution has the thermodynamics of a non interacting ball with size parameter $a = (1 +c_\ell)(d-3)/(\pi(d-2)) $.
The interaction between the gas particles is accommodated by a linear term in the metric function. 

The mass $M$ corresponds to the enthalpy in extended phase space and can be arbitrarily shifted, which was manifest from the $C_M$ term, commensurate with the fact that we can set $C_M=0$ without loss of generality.

Interestingly, the perfect gas black hole satisfies the Weak energy conditions everywhere as long as $c_\ell\leq0$ and $k\neq-1$:
\bea
&&\rho_{perfect} =  \frac{(d-2)(d-3)}{2r^2} - 8\pi c_\ell P, \\
&&(\rho + P)_{perfect} = \frac{d-3}{r^2}.
\eea
In the other cases, there is no generic structure where the WEC are satisfied everywhere.

\section{Case study: why do Reissner-Nordstr\"om black holes approximatively follow the VdW EOS?}
\label{sec:case}

In this section, we will try to understand the qualitative VdW behaviour of the Reissner Nordstr\"om black hole, by comparing its near horizon metric with that of the VdW black hole.

The $AdS$ Reissner-Nordstr\"om black hole solution is given by
\be
f(r) = \frac{r^2}{\ell^2} + 1 - \frac{2M}{r} + \frac{Q^2}{r^2},
\ee
where $Q$ is the charge and $M$ the mass of the black hole. In terms of $A,B$ in \eqref{eq:ansatzf}, it is given by
\be
A = \frac{Q^2}{r^2} - 1,\ B = 0.
\ee

It was shown in \cite{KubiznakMann:2012} that the combination $\frac{P_{\Lambda} v}{T}$ at the critical point, defined as $P_\Lambda'(v_c) = 0 = P_\Lambda''(v_c)$ is the same for the VdW equation of state and the Reissner-Nordstr\"om black hole with the following identification:
\be
a= \frac{3}{4\pi},\ b = 2\sqrt{\frac{2}{3}}Q^2.
\label{eq:abQ}
\ee

For a given black hole to follow a VdW EOS, its metric components and first derivatives of the metric components should be comparable to \eqref{VdWmet}. 
Let us compare explicitly the VdW metric to the Reissner-Nordstr\"om one. First, we fix the constant of integration $C_\ell$ by demanding that the horizon radii are the same, or equivalently that the volume of the black hole is the same:
\be
V_4^Q = V_4^{VdW} \ \Rightarrow C_\ell = -\frac{1}{r_h},
\ee
where $r_h$ is the horizon radius of the Reissner-Nordstr\"om black hole.

Next, we compare the first derivative of the metric coefficient close to the horizon. In the static case, this is equivalent to comparing the temperature of the black holes. 
We find
\be
T_Q = \frac{8 \pi  P_\Lambda r_h^4-Q^2+r_h^2}{4 \pi r_h^3},\ T_{VdW} = -\frac{(b-2 r_h) \left(a+4 P_\Lambda r_h^2\right)}{4 r_h^2},
\ee
where $T_Q,\ T_{VdW}$ are the temperature of the Reissner-Nordstr\"om and Van der Waals black holes respectively. For large horizon radius, the ratio of the temperatures is given by
\be
\frac{T_Q}{T_{VdW}}= 1+\frac{b}{2 r_h}  + \mathcal{O}(r_h)^{-2},
\ee
while for small horizons, 
\be
\frac{T_Q}{T_{VdW}}= \frac{Q^2}{a b \pi r_h}+\frac{2Q^2}{a b^2 \pi}   + \mathcal{O}(r_h)
=1+ \frac{b}{2r_h} + \mathcal{O}(r_h)
\ee
where the latter expression follows upon using \eqref{eq:abQ}.

Matching the two expansions in the intermediate $r_h$ region suggests that the ratio goes like $1+\frac{b}{2 r_h}$ over a wide range of $r_h$. 
As a consequence, the derivatives are qualitatively similar everywhere, except in the limit $r_h\rightarrow0$ (more precisely they deviate in $1/r_h$), for the particular choice of $a,b$ \eqref{eq:abQ}.
% 
% \tcr{\bf [This appears not to be a small-$r$ expansion -- is that correct?]}\tcb{No, it's small $r_h$ expansion. I just wanted to point that the constant term is also 1 for the specific values of a,b.. But I don't think it is very relavant. The relevant region for the comparison is larger $r_h$. I tried to modify the text just below here to clarify.}

\begin{figure}
 \centering
 \includegraphics[scale=.7]{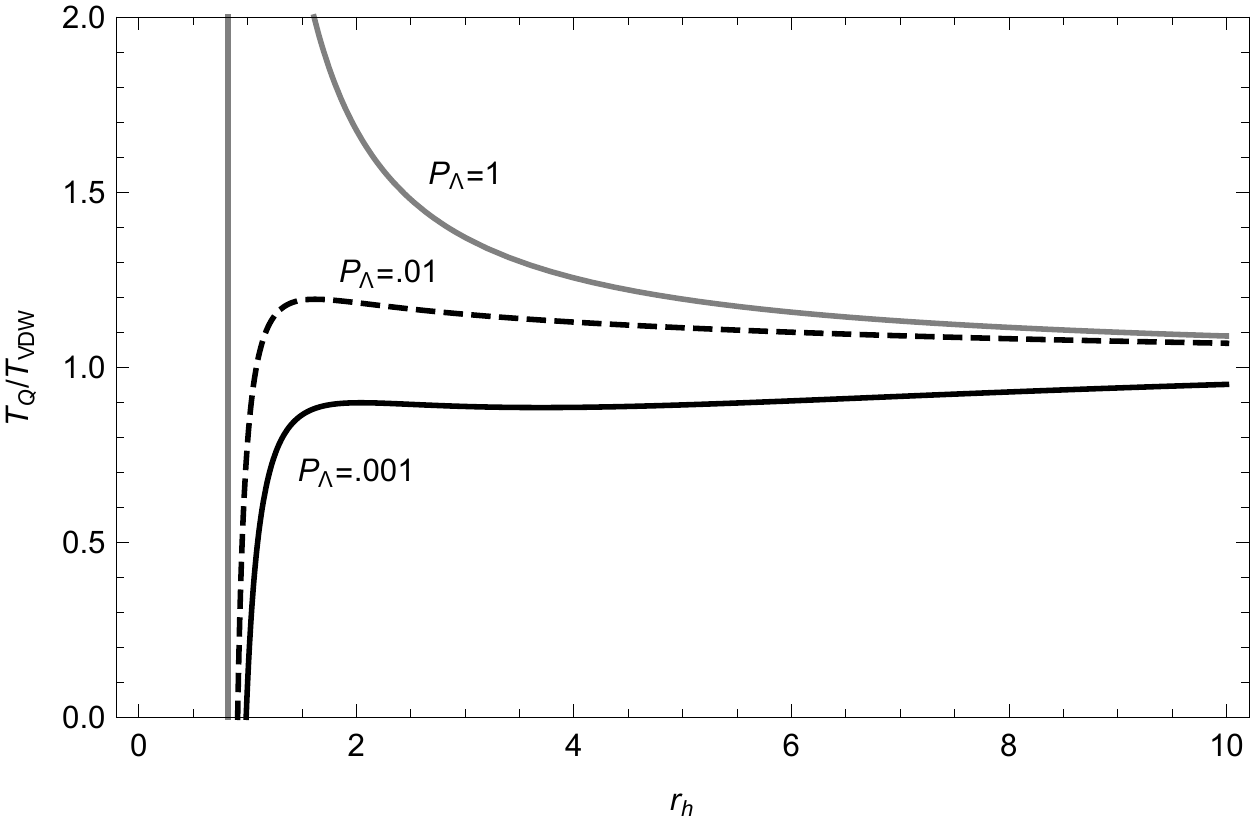}
 \caption{Comparison of the temperature of the Reissner-Nordstr\"om and VDW black holes for different values of $P_\Lambda$.}
 \label{fig:NordVDW}
\end{figure}

In figure \ref{fig:NordVDW}, we show this comparison, and see that it qualitatively agrees for different values of $P_\Lambda$ with a deviation typically for some small horizon radii.

\section{Discussion and conclusion}
\label{sec:concl}
In this paper, we constructed $d$-dimensional $AdS$ black hole metrics whose thermodynamic quantities satisfy the Van der Waals equation of state. These black holes are the exact dual of a Van der Waals fluid in the conformal boundary of the $AdS$ space. Our construction generalizes the solutions presented in \cite{Rajagopal:2014ewa} and is valid for arbitrary numbers of dimensions and spherical, planar and hyperbolic horizons. We discussed in some detail the properties of these black holes and found that they are of positive mass for a certain portion of the parameter space, which we described as a domain of existence.

These black holes are non vacuum solutions. The source to the Einstein equations yielding the Van der Waals black hole has the form of an anisoptropic fluid. The energy associated with the sourcing fluid respects the energy conditions  for a certain range of parameters and in a region sufficiently close to the horizon. We found systematic deviation and violation of the energy conditions far   from the horizon. 

However, this situation can be improved by using a modulating function $\mathcal H$ that is such that $\mathcal H(r_h)=1,\ \mathcal H'(r_h)=0$ and that falls off sufficiently fast. This modulating function can be chosen such that its   defining property is valid over a range of $r_h$ where the energy conditions remains satisfied. This leads in principle to a non vacuum black hole solution that is an exact Van der Waals black hole in some portion of the phase space but otherwise deviates from it outside this portion, yet satisfies the energy conditions everywhere. Indeed, it is impossible in Einstein gravity to devise an exact Van der Waals black hole valid over the whole (extended) phase space that satisfies the energy conditions everywhere. Note however that a small deviation from the weak energy condition can be physically acceptable in $AdS$ spacetime --  for instance the case of negative mass-squared  scalars still leads to a positive $ADM$ mass in AdS \cite{Breitenlohner:1982jf}. 
Despite the systematic violation of the Weak Energy Condition reported here, a stability analysis would be desirable to assess the potential use of the exact VDW black holes in the $AdS/CFT$ picture.

We also analyzed the perfect fluid, interacting point-particle, and non-interacting ball limits of  Van der Waals black holes. In particular, we noted that planar black holes are perfect gas black holes and Tangherlini black holes are non-interacting ball gas black holes. The effect of interaction in the equation of state on the metric is to add a term linear in $r$. These observations may provide useful motivations for a phenomenological description of CFT with the corresponding properties.

Finally, we analyzed the qualitative agreement between the thermodynamics of the Reissner-Nordstr\"om black hole and the Van der Waals black hole by explicitly comparing both metrics in the same coordinate system and horizon location. We found that, as expected, both metrics lead to the same thermodynamic properties for large enough values of the horizon radius.

\section*{Acknowledgement}
This work was supported in part by the Natural Sciences and Engineering Research Council of Canada. We gratefully acknowledge D. Kubiznak for usefull comments and discussion.

\newpage

\appendix
\section{Deriving the $d$-dimensional solution}
In this appendix, we give the details of the derivation of the analytic $d$-dimensional VDW black hole solution.
First, as already pointed out in the text, the solution for the function $B$ is given by 
\be
B = 4\pi r (\mathcal C_\ell r + b).
\ee

Plugging this equation in the equation for $A$ and setting leads to an equation of the form
\be
(r^{d-3}A)' = -\frac{N_0 r^{d-3}(N_1 + D_1 r)}{(D_0+D_1 r)},
\ee
where $ N_1=(d-2)b N_0/(4a\pi),\ D_0=(d-1)bN_0/(4a\pi),\  D_1 = (4+\mathcal C_\ell(d-2)(d-1))N_0/(4 a (d-2)\pi)$.

Denoting $d_0 = D_0/N_0,\ d_1 = D_1/N_0,\ n_1 = N_1/N_0 $, the formal solution to this equation is given by
\bea
&&r^{d-3} A = \mathcal C - \frac{r^d}{d_0^4}\left( \frac{d_1^2}{d} (3n_1 - 2 d_0)+\frac{1}{r^2}\left( \frac{d_0n_1}{d-2}+\frac{d_1(d_0-2n_1)r}{d-1} \right) \right) \nonumber\\
&&\quad- \frac{d_1^3r^{d+1}}{d_0^5(d+1)}\left((2d_0 - 3n_1) _2F_1(1,d+1,d+2;-\frac{d_1}{d_0}r) + (d_0-n_1) _2F_1(2,d+1,d+2;-\frac{d_1}{d_0}r)\right),
\eea
where $_2F_1(a,b,c;x)$ is the Hypergeometric function. For integer values of $d$, the solution reduces to a finite polynomial with logarithmic terms, and $\mathcal C$ is an integration constant.

%%%%%%%%%%%%%%%%%%%%%%%%%%%%%%%%%%%%%%%%%%%%%%%%%%%%%%%%%%%%%%%%%%%
\clearpage
 \bibliographystyle{h-physrev4}
 \bibliography{vdwbh}
%%%%%%%%%%%%%%%%%%%%%%%%%%%%%%%%%%%%%%%%%%%%%%%%%%%%%%%%%%%%%%%%%%%
%%%%%%%%%%%%%%%%%%%%%%%%%%%%%%%%%%%%%%%%%%%%%%%%%%%%%%%%%%%%%%%%%%%
\end{document}